\documentclass[12pt,preprint]{aastex}

\begin{document}

\title{Possible Implications of Relatively High Levels of Initial $^{60}$Fe in Iron Meteorites
for the Non-Carbonaceous -- Carbonaceous Meteorite Dichotomy and Solar Nebula Formation}

\author{Alan P. Boss}
\affil{Earth \& Planets Laboratory, Carnegie Institution
for Science, 5241 Broad Branch Road, NW, Washington, DC 20015-1305}
\authoremail{aboss@carnegiescience.edu}

\begin{abstract}

 Cook et al. (2021) found that iron meteorites have an initial abundance ratio of
the short-lived isotope $^{60}$Fe to the stable isotope $^{56}$Fe of 
$^{60}$Fe/$^{56}$Fe $\sim$ $(6.4 \pm 2.0) \times 10^{-7}$. This appears to require
the injection of live $^{60}$Fe from a Type II supernova (SN II) into the presolar
molecular cloud core, as the observed ratio is over a factor of ten times
higher than would be expected to be found in the ambient interstellar medium (ISM)
as a result of galactic chemical evolution. The supernova triggering and injection
scenario offers a ready explanation for an elevated initial $^{60}$Fe level, and in
addition provides a physical mechanism for explaining the non-carbonaceous -- 
carbonaceous (NC-CC) dichotomy of meteorites. 
The NC-CC scenario hypothesizes the solar nebula first
accreted material that was enriched in supernova-derived nuclides, and
then later accreted material depleted in supernova-derived nuclides. While the
NC-CC dichotomy refers to stable nuclides, not short-lived isotopes like $^{60}$Fe,
the SN II triggering hypothesis provides an explanation for the otherwise
unexplained change in nuclides being accreted by the solar nebula.
Three dimensional hydrodynamical models of SN II shock-triggered collapse show
that after triggering collapse of the presolar cloud core, the shock front sweeps away
the local ISM while accelerating the resulting protostar/disk to a speed
of several km/s, sufficient for the protostar/disk system to encounter within 
$\sim$ 1 Myr the more distant regions of a giant molecular cloud complex that
might be expected to have a depleted inventory of supernova-derived nuclides.
 
\end{abstract}


\section{Introduction}
 
  Laboratory studies of the initial abundances of the short-lived isotope 
$^{60}$Fe (half-life of 2.62 Myr) have produced values differing by orders of
magnitude. Tachibana et al. (2006) determined an initial ratio 
of $^{60}$Fe/$^{56}$Fe $\sim 5 - 10 \times 10^{-7}$ in their studies of
high Fe/Ni ferromagnesian chondrules from the ordinary chondrites (OC)
Semarkona and Bishunpur. Tang \& Dauphas (2012) performed whole
rock analyses of a range of meteorites, including unequilibrated ordinary
chondrites (UOC), and found a much lower initial abundance ratio,  
$^{60}$Fe/$^{56}$Fe $\sim 1.15 \times 10^{-8}$, a level they considered 
to be representative of the general ISM. However, when Mishra \& Goswami 
(2014) examined seven chondrules from UOC meteorites, they
found an initial ratio of $^{60}$Fe/$^{56}$Fe $\sim 7 \times 10^{-7}$.
Mishra \& Chaussidon (2014) studied three chondrules from the OC
Semarkona and the carbonaceous chondrite Efremovka, finding initial
ratios of $^{60}$Fe/$^{56}$Fe in the range of $\sim 2 - 8 \times 10^{-7}$.
Mishra et al. (2016) inferred ratios of $\sim 8 - 11 \times 10^{-7}$ for
chondrules from an UOC, while Telus et al. (2018) found initial ratios
in the range of $\sim 0.5 - 3 \times 10^{-7}$ in other UOC chondrules.

 Telus et al. (2016) attributed the differences between whole rock
analyses and {\it in situ} studies of individual chondrules to aqueous 
alteration along chondrule fracture lines, either on the parent body or 
prior to recovery on the Earth, which could skew the whole rock 
analyses toward lower initial $^{60}$Fe/$^{56}$Fe ratios.
Trappitsch et al. (2018) reanalyzed a Semarkona chondrule  with
a different {\it in situ} technique, finding a low initial ratio of 
$^{60}$Fe/$^{56}$Fe $\sim (3.8 \pm 6.9) \times 10^{-8}$ that is 
consistent with the low end of the range found by some other 
{\it in situ} analyses and with the low values found by the 
whole rock measurements of Tang \& Dauphas (2012).
The Trappitsch et al. (2018) results allow a solar system initial
ratio as high as $1.07 \times 10^{-7}$.

 The Tang \& Dauphas (2012) result has been accepted by some 
as evidence of meteoritic $^{60}$Fe being the result of galactic 
chemical evolution (e.g., Forbes et al. 2021). Others have been 
more circumspect. Vescovi et al. (2018) simply concluded that the 
initial $^{60}$Fe/$^{56}$Fe ratio lies between $10^{-8}$ and $10^{-6}$,
while Lugaro et al. (2018) decided that the initial $^{60}$Fe/$^{56}$Fe 
ratio has not been determined well enough to draw any conclusions
about the source of this short-lived isotope.

 Given this checkered past, the Cook et al. (2021) study represents a
potentially transformational approach to determining the initial
$^{60}$Fe/$^{56}$Fe ratio. Rather than search for Fe-Ni-rich phases
in the primarily silicate mineralogy of chondritic meteorites, 
Cook et al. (2021) analyzed 13 samples from two groups of 
magmatic iron meteorites, the common octahedrites (group IID) and
the rare ataxites (group IVB), the former with average to high Ni
contents, the latter with very high Ni. Magmatic iron meteorites
offer the advantage of samples of a well-mixed iron melt, sidestepping
the concerns about whole rock versus {\it in situ} measurements that
characterize chondritic meteorites. The study was able to rule out 
contamination of the Fe and Ni isotopes by galactic cosmic-rays.
Using the common assumption that the iron meteorites formed from
the chondritic reservoir, Cook et al. (2021) found that an initial ratio
of $^{60}$Fe/$^{56}$Fe $\sim (6.4 \pm 2.0) \times 10^{-7}$ characterized 
their samples of these two groups of iron meteorites, a value considerably 
higher than expected as a result of galactic chemical evolution (GCE; 
e.g., Huss et al. 2009). Cook et al. (2021)  concluded that the solar system
must have been injected with additional live $^{60}$Fe, well above the
level to be expected from GCE, and noted that deriving the injected
additional $^{60}$Fe from a SN II also best explained the deficits in
$^{60}$Ni and $^{56}$Fe abundances they measured in their sample 
of iron meteorites.

 If correct, the Cook et al. (2021) study appears to settle the brouhaha
over the initial abundance of $^{60}$Fe in the solar nebula, and thus
provides meteoritical evidence in support of the supernova triggering 
and injection hypothesis that was first proposed by Cameron \& Truran (1977).
Lee et al. (1976) found evidence of live initial $^{26}$Al (half-life of 0.72 
Myr) in Ca, Al-rich refractory inclusions (CAIs) from the Allende meteorite,
leading Cameron \& Truran (1977) to hypothesize nucleosynthesis
of $^{26}$Al in a SN II and rapid incorporation of the $^{26}$Al into CAIs
following injection into the presolar cloud by the SN II shock wave. 
$^{26}$Al is also synthesized during the Wolf-Rayet (WR) phase 
of massive stars, prior to a core-collapse supernova explosion,
and mixed into the interstellar
medium by the WR star outflow, and so the solar system level of 
$^{26}$Al has been argued to be a result of GCE (e.g., Young 2016; 
Reiter 2020) or of triggering by a WR outflow (Dwarkadas et al. 2017).
However, $^{60}$Fe is not produced in significant amounts by WR stars,
but it as well as $^{26}$Al is readily produced by SN II (e.g., Tur et al. 2010),
making the initial $^{60}$Fe abundance the key meteoritical test for a 
SN II being the primary source of the live $^{60}$Fe. 
Protosolar cosmic rays have also been suggested as a source of $^{26}$Al
(most recently by Gaches et al. 2020) but such cosmic rays fail to 
produce the required $^{60}$Fe.

 A long series of papers, from Foster \& Boss (1996) through to
Boss (2019), has used detailed multidimensional hydrodynamics codes
to show that suitable SN II shock waves are able to trigger the 
self-gravitational collapse of a molecular cloud core, while simultaneously
injecting the short-lived isotopes into the collapsing cloud through the
Rayleigh-Taylor instability at the shock-cloud boundary. These papers
have proven the viability of the Cameron \& Truran (1977) hypothesis
for the origin of the short-lived isotopes inferred from meteoritical
analyses such as those of Cook et al. (2021). 

 The purpose of this paper is to learn whether this series of shock-triggered 
collapse models might have additional implications for explaining the 
non-carbonaceous -- carbonaceous (NC-CC) dichotomy of meteorites.
Warren (2011) was the first to show that meteoritical Cr, Ti, and O 
stable isotope abundances fall into two distinct groups, the NC and CC.
Recent work by Nanne et al. (2019) and Lichtenberg et al. (2021)
has developed models for creating and preserving the NC-CC
dichotomy. Both scenarios hypothesize that the solar nebula first
accreted material that was enriched in supernova-derived nuclides, and
then later accreted material depleted in supernova-derived nuclides.
No explanation is offered for a physical mechanism to explain this
significant difference in accreted matter other than isotopic heterogeneity
in the local ISM and an extended period of accretion from the ISM.
Thus we seek here to find a more dynamically-based explanation for the 
NC-CC dichotomy. We extend the preferred triggered-collapse model of 
Boss (2019) to a much larger calculational volume and use it to track the 
evolution of matter initially in four distinct initial regions to learn the effect 
of shock-wave triggering on the time evolution of matter accreted by the 
presolar cloud and nebula.

\section{Numerical Hydrodynamics Code}

 The new model presented here was calculated in the same manner as
the suite of shock-triggered collapse and injection models of Boss (2019).
The three-dimensional hydrodynamics code used was once again FLASH 4.3,
based on the algorithms developed by Fryxell et al. (2000). The FLASH
codes are adaptive-mesh refinement (AMR) codes, ideal for following the
sharp gradients in density and temperature associated with shock fronts
as they traverse more uniformly varying regions of the ISM, such as
molecular cloud cores. In addition to the AMR feature of FLASH, the 
models use a  sink particle (as developed by Federrath et al. 2010) to 
represent the newly formed, high density protostar as a point source of
gravity able to accrete matter from its surroundings. As in Boss (2019),
the new model uses the FLASH 4.3 multigrid Poisson solver, and as
a result uses a Cartesian coordinate grid with one top grid block
with eight top grid cells in each coordinate direction. The model begins
with a maximum of six levels of refinement on the top grid block.
Because the computational volume is a rectangular cuboid with sides 
of length $4 \times 10^{17}$ cm in $\hat x$ and $\hat z$ and 
$1.12 \times 10^{18}$ cm in $\hat y$, the resulting cells are not cubical, 
being almost three times as long in the $\hat y$ direction as in $\hat x$ 
and $\hat z$. [Note that in Boss (2019), the rectangular cuboid had 
a significantly smaller length in $\hat y$ of $8.2 \times 10^{17}$ cm.]
A single increase in the level of refinement means that each of the
three sides of the computational cells are halved in length, leading to the 
increased spatial resolution that is needed to be able to accurately 
follow the dynamics of shock-triggered protostar collapse.
With six levels of refinement, the smallest cell size in $\hat x$ and 
$\hat z$ is $4 \times 10^{17}/(2^5 \times 8) = 1.6 \times 10^{15}$ cm
and a size 2.8 times larger in $\hat y$. When the refinement level
is increased to seven, the smallest cell size is $0.8 \times 10^{15}$ cm
in $\hat x$ and $\hat z$.
 
\section{Initial Conditions}

 The model presented here is identical to one of the models (O) published
in Boss (2019), with two exceptions. First, the length of the calculation box 
was extended in the direction of propagation of the shock front in order
to learn more about the effect of the shock front on the ISM gas and
dust downstream from the target molecular cloud core. Second, several
new color fields were defined, in order to better define the fate of the
surrounding ISM matter. The color fields are defined as being 
initially nonzero in specific regions of the initial configuration, such
as inside the target cloud or inside the shock front, and these fields
thereafter evolve and trace the location and density of this material 
as the calculation proceeds.
The models in Boss (2019) focused on the material initially inside
the shock front, in order to estimate the SN II injection efficiency, 
and did not specify color fields that tracked the material initially in 
the target cloud or the surrounding ISM. As a result, the Boss (2019)
models could not separate the evolution of the initial target cloud
material from that of the surrounding ambient ISM, which is the key
goal of this paper.

 As in Boss (2019), the initial conditions consist 
of a stationary target molecular cloud core that is about to be struck by 
a planar shock wave (Figure 1). The target cloud core and the 
surrounding gas are initially isothermal at 10 K, while the shock front 
and post-shock gas are isothermal at 1000 K. The target cloud consists
of a spherical cloud core with a radius of of 0.053 pc and a Bonnor-Ebert 
radial density profile. The central density is chosen to produce an
initial cloud with a mass of 3.04 $M_\odot$, embedded in a background 
rectangular cuboid of gas with a mass of 1.63 $M_\odot$ 
and with random noise in the background density 
distribution. The cloud core is assumed to be in solid body rotation 
about the direction of propagation of the shock wave (the $- \hat y$
direction) at an angular frequency $\Omega_i = 3 \times 10^{-14}$ rad s$^{-1}$.
The initial shock wave has a speed of 40 km s$^{-1}$, a width of 
$3 \times 10^{-4}$ pc, and a density of $7.2 \times 10^{-18}$ g cm$^{-3}$.
These choices are based on previous modeling, which examined a 
wide range of shock parameters (Boss et al. 2010; Boss \& Keiser 2010), 
in order to choose shock parameters suitable for triggered collapse
and injection of SN II-derived short-lived isotopes.

\section{Results}

 Figures 1 though 5 shows the initial configuration and the four different 
color fields used to trace the evolution of different regions in the model.
The first color field (denoted mass scalar 1 or ms1, in the FLASH code)
depicts the matter in the target cloud, while ms2 traces the matter in
the shock front, ms3 does the same for the matter behind the shock front,
and ms4 follows the ambient ISM surrounding the target cloud core.

 The time evolution of this model proceeds exactly the same as the
corresponding model O in Boss (2019): the shock front smacks the
top edge of the target cloud core, leading to a Rayleigh-Taylor instability
at the shock-cloud interface. This instability allows shock front material
carrying SN II-produced short-lived isotopes
such as $^{60}$Fe and $^{26}$Al to be injected into the
target cloud core, which is soon compressed sufficiently by the shock
front to initiate sustained, self-gravitational collapse. Once the collapsing
region exceeds a critical density of $10^{-15}$ g cm$^{-3}$ (0.048 Myr),
a sink cell is formed at the location of the density maximum, and this
sink cell thereafter accretes the gas and dust in its vicinity, using the
same sink cell parameters as used in model O in Boss (2019). 
The model started with a maximum of six levels of refinement, which 
was increased to seven levels after 0.050 Myr of evolution.
The portions of the shock that do not strike the target cloud
exit off the calculational grid by 0.010 Myr.
 
 Figures 6 through 10 depict the results for the density and four color
fields at the final time calculated for the model of 0.063 Myr. This 
new AMR model required a run time of three months on three 32-core 
nodes of the Carnegie memex cluster. At the final time, the sink cell 
is located close to the center of the density maximum evident in 
Figure 6 (at $y = - 1.32 \times 10^{17}$ cm), has been
accelerated by the shock front to a speed of 3.2 km/s in the
direction of the initial shock front, and has acquired a mass of
$\sim 0.5 M_\odot$. The protostar is accreting mass at a rate
of $\sim 10^{-6} M_\odot$/yr, implying that it will grow to a final
mass of $\sim 1 M_\odot$ in $\sim 0.5$ Myr if that accretion
rate could be sustained indefinitely. However, at the final time,
the mass accretion rate is in decline, as the nearby gas and dust
available for accretion is being depleted. 

 Figure 7 shows that the initial cloud core
has been only partially triggered into self-gravitational collapse,
with the majority of the initial mass of 3.04 $M_\odot$ having
been accelerated beyond the gravitational reach of the 
accreting protostar formed by shock wave compression.
Figure 8 indicates that, as in the previous model O (Boss 2019),
the shock front material has been mixed into the same region 
occupied by the initial cloud core gas and dust, leading to an
injection efficiency into the collapsing protostar essentially
the same as was previously determined, as discussed in
detail in Boss (2017). Figure 9 shows that the low-density,
hot post-shock gas follows right behind the shock front
matter with only minimal mixing into the compressed region
of high density gas and dust.

 Figure 10 presents the key result of this simulation: the material
initially in the immediate vicinity of the target cloud core has been 
quite efficiently swept away from the region of the protostar and 
of the material that it is still accreting. Only an insignificant
amount of the initial ambient gas and dust remains close enough
to the protostar to be accreted, less than 1/20 (by mass) 
compared to the shock front matter at densities above
$2 \times 10^{-16}$ g cm$^{-3}$. The model shows that SN II-derived
isotopes will be injected thereafter, with only minimal subsequent
accretion of isotopes from the initial ambient ISM in the immediate
vicinity of the target molecular cloud core, which is assumed
to be initially pristine and unpolluted by SN II-shock ejecta. 
This sweeping away of the local ISM allows the protostar to
experience two distinct phases of accretion, first due to SN II-shock
injection, and later during its subsequent traverse of the GMC.
Figures 11 and 12 provide close-in views of the model at the final
evolution time, showing the newly-formed protostar/protoplanet 
disk surrounding the location of the protostar (i.e., sink cell).
Figure 12 shows that only faint wisps of the initial ambient ISM
material still remain in the region of the protostar.
Figure 13 shows that the shock wave isotopes have been
injected deep within the protostar and disk system, as a result
of the Rayleigh-Taylor fingers that pierce the target cloud core
early in the evolution. Finally, Figure 14 depicts the temperature
distribution at the final time, showing that the protostar is located
well behind the high temperature (1000 K) shocked region and
retains its initial temperature of 10 K as a result of molecular
line cooling in optically thin regions (Boss et al. 2008). The 
protostar thereafter continues its evolution into the GMC
with its ambient temperature of 10 K. 

 Figure 11 superficially resembles Figure 5 of Boss (2019),
which depicts the results of a model (N) with an initial cloud
rotation rate 2/3 of that in the current model, based on model O and
shown in Figure 6 of Boss (2019). In Boss (2019), model O produced 
a large-scale disk by 0.121 Myr, with a diameter of order 1000 au,
whereas model N did not produce a large-scale disk  by the
time that the model was halted, 0.081 Myr. The present model
shown in Figure 11 was halted at 0.063 Myr, even earlier than
model N in Boss (2019), as the result of an increasingly smaller 
time step problem that did not afflict model O in Boss (2019).
As the present model is identical to model O, save for the extension
of the numerical grid from $y = - 2 \times 10^{17}$ cm to
$y = - 5 \times 10^{17}$ cm effected in order to better study the fate
of the surrounding ISM matter, one must ascribe this small time step 
problem to the computational cells being increasingly elongated in $\hat y$ 
that had to be employed in order to use the improved multigrid
Poisson solver of FLASH 4.3, as explained by Boss (2019).
Hence one can expect that the present model would produce a
large-scale disk identical to that of model O if the model could
be calculated as far as model O in Figure 6 of Boss (2019), i.e.,
to 0.121 Myr. A careful examination of the present model in Figure 11
shows that a disk is beginning to form, as there are distinct "horns"
of accreting matter both above and below the location of the central
protostar, with the horns above the protostar being clearly bent
outward compared to those of model N in Figure 5 of Boss (2019),
indicative of the higher initial rotation rate of the present model compared
to model N and of the subsequent disk formation in model O. 
Regardless of the small time step problem, the present model's
extension in $\hat y$ allows Figure 10 to demonstrate that the shock
front effectively clears away the residual ISM matter from the 
vicinity of the protostar and disk system, a fact that cannot be
gleaned from model O's Figure 6 in Boss (2019), where the disk system 
is about to reach the end of the numerical grid.

 The mass of the initial shock front is 0.50 $M_\odot$ and it is
traveling at a speed of 40 km s$^{-1}$, giving it a total momentum
of 20 $M_\odot$ km s$^{-1}$. At the final time, the protostar has 
a mass of $\approx 0.5 M_\odot$ and has been accelerated to 
3.2  km s$^{-1}$ in the direction of the initial shock front, giving it a total
momentum of $\approx 1.6 M_\odot$ km s$^{-1}$. Thus the protostar
has acquired less than 1/10 of the total initial momentum of 
the shock front. At the rate of  3.2  km s$^{-1}$,
the protostar/disk system will traverse $\sim$ 3 pc through 
the background giant molecular cloud complex
(GMC) in 1 Myr, accreting gas and dust from 
other, more distant regions of the GMC that have not been
polluted recently by a SN II explosion. Given that GMC diameters
range from $\sim$ 5 pc to $\sim$ 200 pc, there would appear
to be an adequate GMC volume for a protostar launched at $\sim$ 3 km/s
to scatter off other protostars in the GMC star-forming regions
and accrete further ISM material and non-SN II derived
isotopes.

\section{Implications for the NC-CC Dichotomy}

 By this time, the implications of this model should be clear: they 
provide a physically reasonable explanation for the NC-CC dichotomy,
as advanced by Nanne et al. (2019, see their Figure 7) and 
Lichtenberg et al. (2021, see their Figure 6). These authors
hypothesized that the solar nebula initially accreted material that 
was enriched in SN II-derived isotopes, and some time
later accreted material depleted in SN II-derived isotopes.

 While no time scale for these two accretion phases is presented
by either Nanne et al. (2019) or Lichtenberg et al. (2021) other than 
the phrases "early infall" and "late infall", the present model suggests
that the "early infall" phase lasted for $\sim 0.1$ Myr, based on the
final model time reached, while the "late infall" phase lasted for $\sim 1$ Myr, 
based on the speed of the protostar and GMC sizes. These suggestions
are supported by the work on the NC-CC dichotomy by Kruijer et al.
(2020), whose Figure 5 explicitly states that the "early infall" occurs
at "t = 0" Myr, implying a time scale much less than 1 Myr, along
with assuming the simultaneous formation of CAIs, while
"late infall"  stops by 1 Myr, when Jupiter is supposed to have formed
and prohibited mixing of the inner NC and outer CC  reservoirs.

Isotopic heterogeneity throughout the GMC where the solar system 
was formed is required, but the present model provides a physical reason
for the change in the SN II-derived isotopic composition of the matter 
being accreted by the protosun and solar nebula. Jacquet et al.
(2019) make a similar argument for isotopic heterogeneity in
the matter accreted by the solar nebula without specifying a
physical mechanism to explain the origin of the heterogeneity.
Others have argued that physicochemical processing of dust
grains could be a better explanation for certain isotopic anomalies
($^{150}$Nd) than heterogeneous infall (Saji et al. 2021).

Hopp et al. (2022) have argued that Fe isotopic abundances in
iron meteorites reflect the same NC-CC dichotomy as other
meteorites. They conclude that the Fe isotope dichotomy can be
explained by nuclear statistical equilibrium in either type Ia SN or
in core-collapse SN, i.e., SN II.

 Ideally the present calculation could be extended indefinitely
further in time to capture both the "early infall" and "late infall"
phases advanced by Nanne et al. (2019) and Lichtenberg et al.
(2021). This is not possible for the present model for two reasons.
First, increasingly smaller time steps are required to push the
calculation any farther in time than the 0.063 Myr shown in
Figures 6 through 12, effectively halting this particular model.
Second, even without the time step problem, the protostar and 
disk system seen in Figure 12 at $y = - 1.3 \times 10^{17}$ cm
would reach the end of the computational volume at
$y = - 5 \times 10^{17}$ cm within about another 0.038 My,
traveling at 3.2 km/s, too quickly to transition to the "late infall"
phase. Note that prior to reaching the bottom of the computational
volume, the dense gas that is accreting onto the central system
(orange-red colors in Figure 12) will have been accreted, as the
free fall time for collapse at a gas density of $10^{-16}$ g cm$^{-3}$
is 0.0067 Myr, effectively ending the "early infall" phase.

 If the present scenario is deemed interesting, future work would
be needed to consider a more global model of an entire GMC struck
by a SN II shock wave that follows the progression of the resulting
shock-triggered protostar(s) as they traverse the GMC. In lieu of
such an ambitious three-dimensional hydrodynamics model, we can
predict the mass accretion rate that should characterize the "late infall"
phase. The protostar will accrete gas from the region of the GMC
it traverses at a rate that can be calculated by the 
Bondi-Hoyle-Lyttleton (BHL) formula given by Ruffert \& Arnett (1994):
$\dot{M}_{BHL} = \pi R_A^2 \rho_A v_A$, where the accretion radius
$R_A$ is given by $R_A = 2 G M / v_A^2$, with $G$ being the 
gravitational constant, $\rho_A$ being the ambient GMC gas density,
and $v_A$ is the speed of the protostar with respect to the ambient gas. 
With an ambient GMC gas density of $\rho_A = 10^{-21}$ g cm$^{-3}$, 
typical of GMC average densities and the same as the background
gas in the present model (see Figure 1), $v_A = 3.2$ km/s, and
a protostar and disk system mass of $1 M_\odot$, we obtain a mass
accretion rate of $\dot{M}_{BHL} = 10^{-4} M_\odot$/Myr of ambient
GMC gas and dust. If the protostar system should pass through a
dense molecular cloud core with a mean density of 
$\sim 10^{-19}$ g cm$^{-3}$, as assumed in the present model
(Figure 1), then the mass accretion rate would increase to
$\dot{M}_{BHL} \approx 10^{-2} M_\odot$/Myr while traversing the
molecular cloud core, which would take about 0.03 Myr at 3.2 km/s.
These estimates suggest that the mass of pre-existing GMC gas and 
dust that was not enriched by the triggering SN II shock wave
but was accreted by the protostar during a "late infall" phase
lasting $\sim$ 1 Myr would be in the range of $10^{-4} M_\odot$
of ambient matter to $3 \times 10^{-4} M_\odot$ of dense cloud core
gas and dust. If the "late infall" phase lasts longer than 1 Myr,
given that a protostar moving at 3.2 km/s travels 3 pc in 1 Myr,
but GMCs can span $\sim$ 5 pc to $\sim 200$ pc in size, the 
protostar could continue to accrete ambient GMC matter for longer 
than 1 Myr. Either way, the total accreted mass in the "late infall"
phase is likely to be of order $10^{-4} M_\odot$
to $\sim 3 \times 10^{-4} M_\odot$.

 What are the implications of these estimates for the masses of
the CC and NC component? Given that model O in Boss (2019),
the model recalculated here with a longer computational volume,
produced a disk with an initial mass of $0.05 M_\odot$, representing
the "early infall" or CC component, whereas the "late infall" or NC
component mass accretion estimate is  in the range of
$10^{-4} M_\odot$ to $\sim 3 \times 10^{-4} M_\odot$, this implies 
that the CC component was about 170 to 500 times as massive
as the NC component. While specific values for the masses of the
"early infall" enriched matter and "late infall" depleted matter are
not given by Nanne et al. (2019), Lichtenberg et al. (2021)
suggest that their Reservoir I (NC) had a total mass of about
a Earth mass of planetesimals, while their Reservoir II (CC)
had about a Jupiter mass of planetesimals, implying that the
"early infall" CC matter was about 318 times as massive as the
"late infall" NC matter. Given the uncertainties involved in
both the present model and the Lichtenberg et al. (2021) estimates,
having these two estimates agree to within a factor of two is
remarkable and suggests that the scenario proposed herein is
worthy of further scrutiny.
  
\section{Conclusions}

 Observations of star-forming regions have demonstrated that
star formation can be triggered by supernova explosions 
(e.g., Bialy et al. 2021). The meteoritical evidence discussed
in this paper, coupled with the results of the new model
presented here, along with the previous papers in this series,
appear to provide a reasonable argument that the solar system
was formed as a result of the interaction of a SN II shock wave 
with a dense molecular cloud core residing within a GMC complex.

\acknowledgments

 The computations were performed on the Carnegie Institution memex computer
cluster (hpc.carnegiescience.edu) with the support of the Carnegie Scientific
Computing Committee. I thank Floyd Fayton for his invaluable assistance with 
the use of memex and Myriam Telus for discussions about the Cook et al. (2021) 
paper. The referees provided numerous useful suggestions for improving
the paper. The software used in this work was in large part developed 
by the DOE-supported ASC/Alliances Center for Astrophysical Thermonuclear 
Flashes at the University of Chicago.

\begin{figure}
\vspace{-1.0in}
\includegraphics[scale=.80,angle=0]{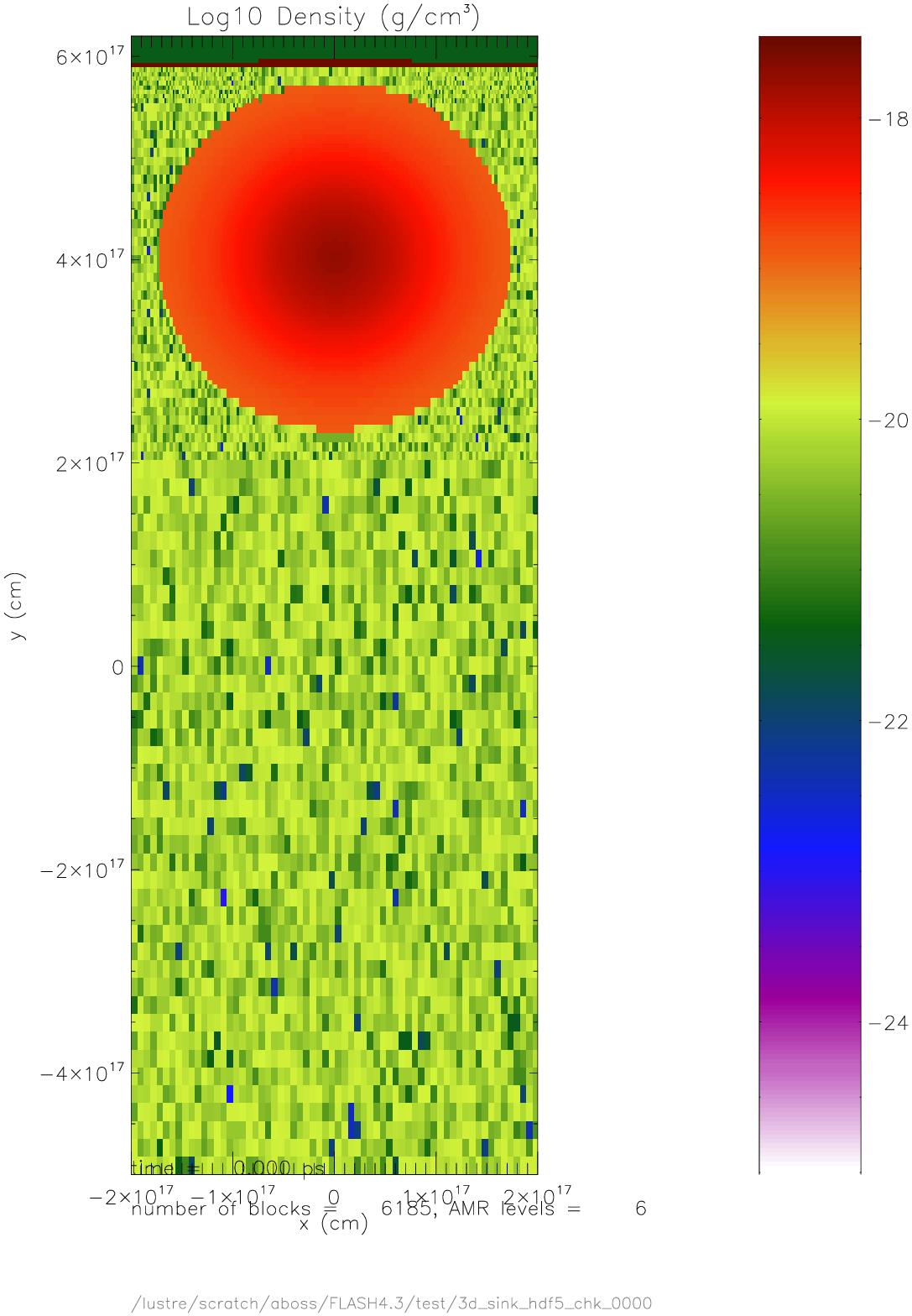}
\vspace{0.5in}
\caption{Initial log density cross-section ($z$ = 0) showing the entire computational 
grid with a maximum of six levels of refinement. }
\end{figure}
\clearpage

\begin{figure}
\vspace{-1.0in}
\includegraphics[scale=.80,angle=0]{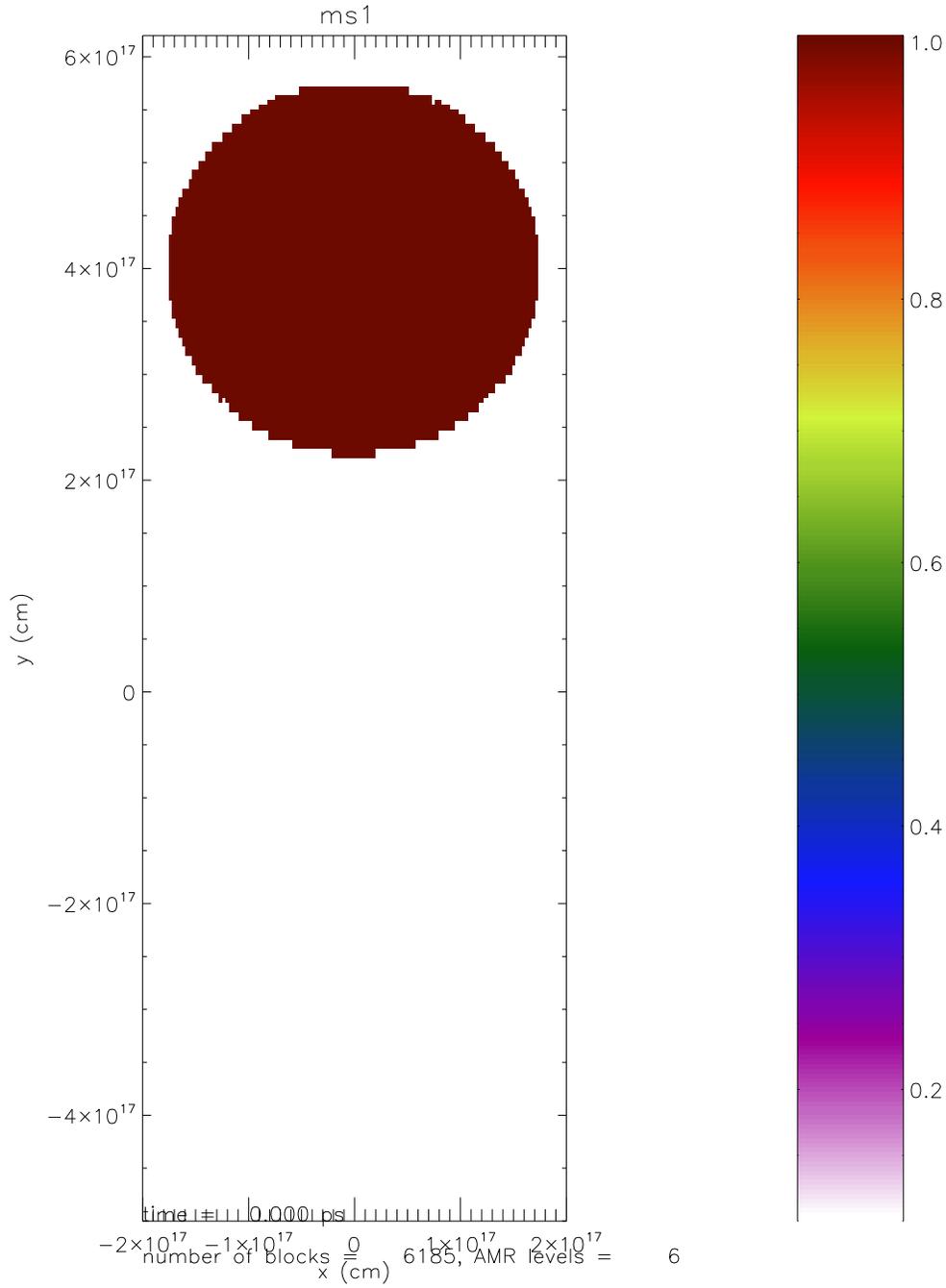}
\vspace{0.5in}
\caption{Initial cross-section ($z$ = 0) of color field 1, representing the material
initially within the target cloud core, plotted as in Figure 1. The target cloud is 
initially stationary.}
\end{figure}
\clearpage

\begin{figure}
\vspace{-1.0in}
\includegraphics[scale=.80,angle=0]{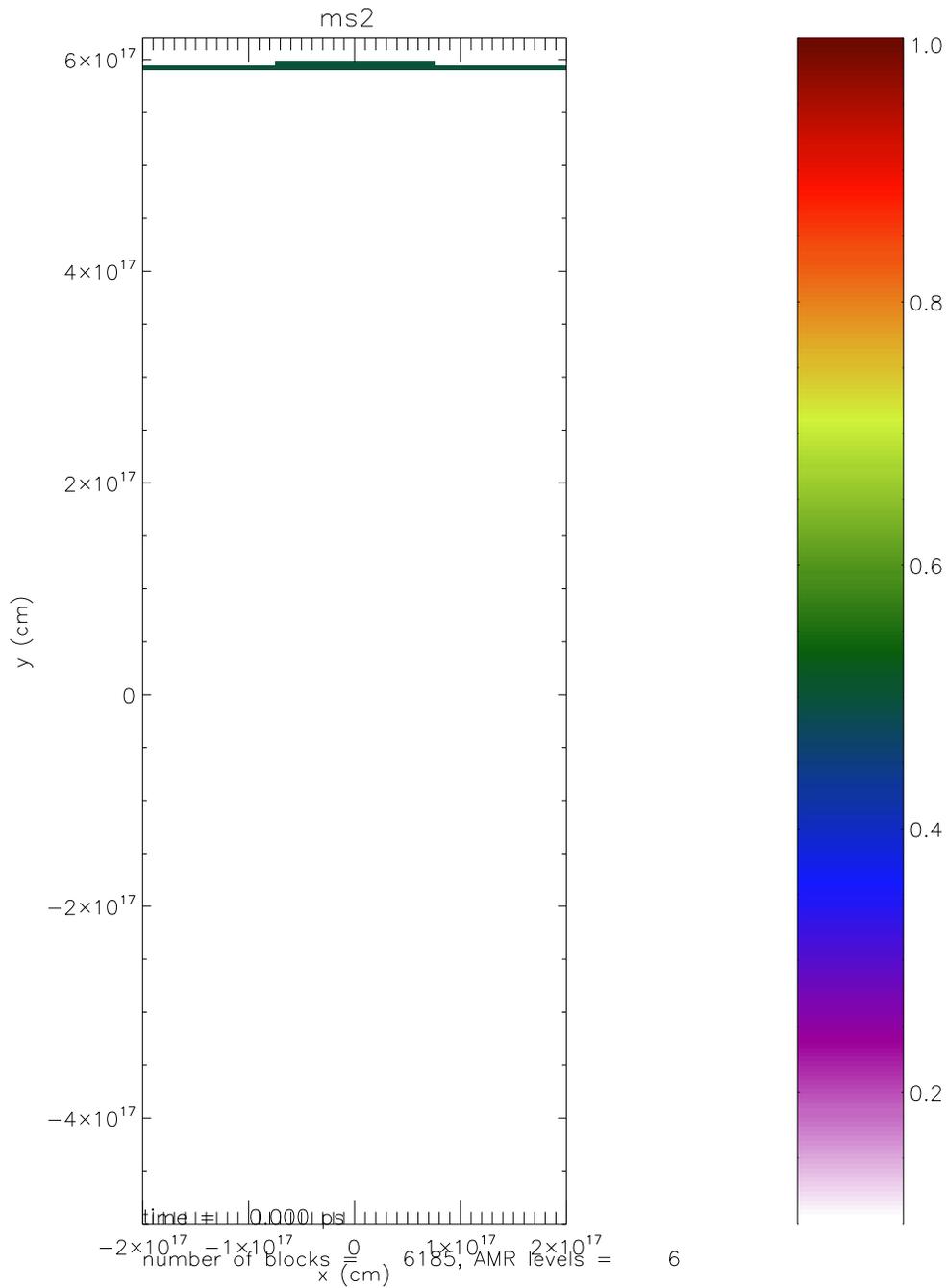}
\vspace{0.5in}
\caption{Initial cross-section ($z$ = 0) of color field 2, representing the material
initially within the shock front, plotted as in Figure 1. The shock is initially
moving downwards at 40 km $^{-1}$.}
\end{figure}
\clearpage

\begin{figure}
\vspace{-1.0in}
\includegraphics[scale=.80,angle=0]{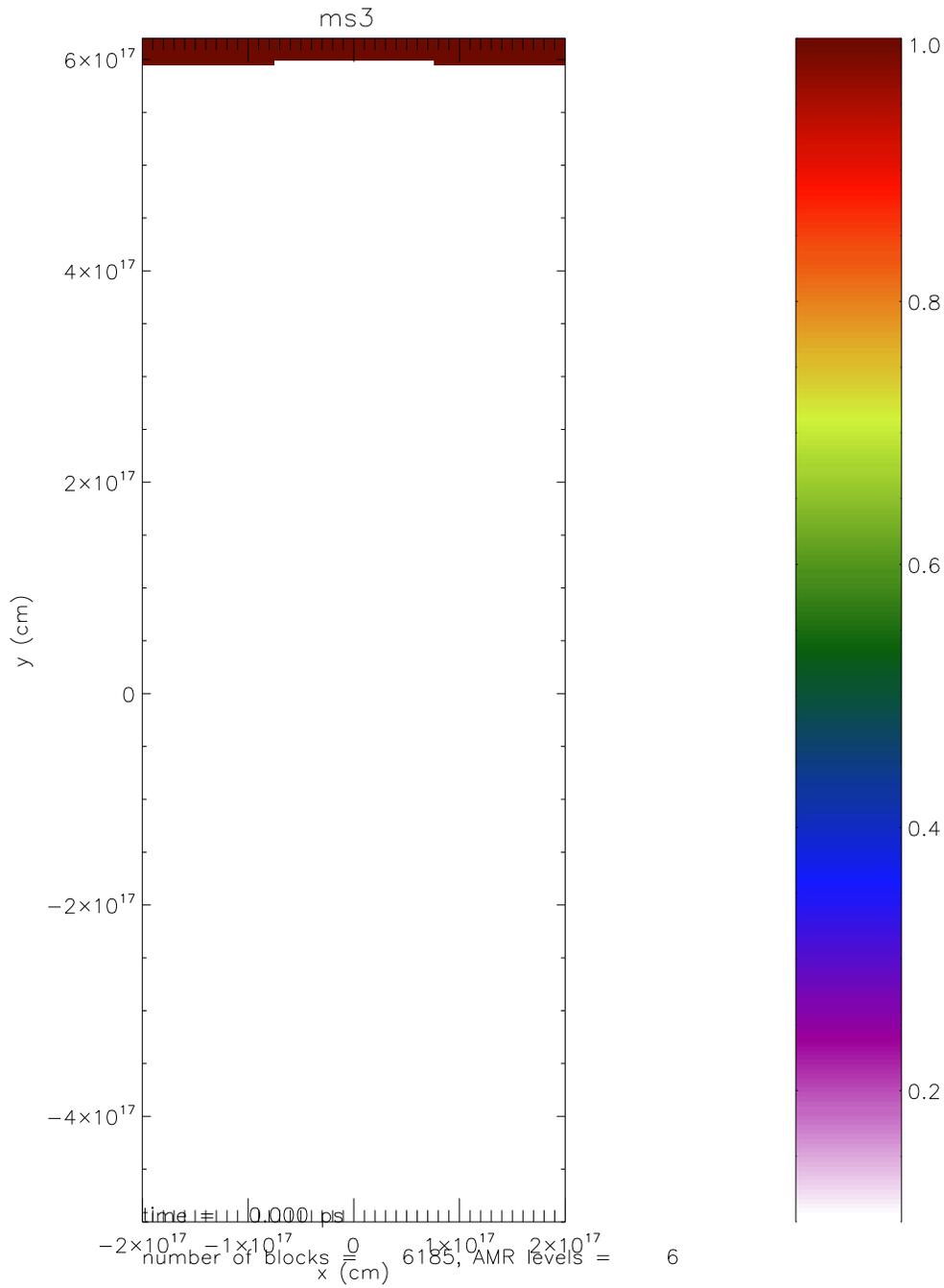}
\vspace{0.5in}
\caption{Initial cross-section ($z$ = 0) of color field 3, representing the material
initially behind the shock front, plotted as in Figure 1. The material behind
the shock is also moving downwards at 40 km $^{-1}$.}
\end{figure}
\clearpage

\begin{figure}
\vspace{-1.0in}
\includegraphics[scale=.80,angle=0]{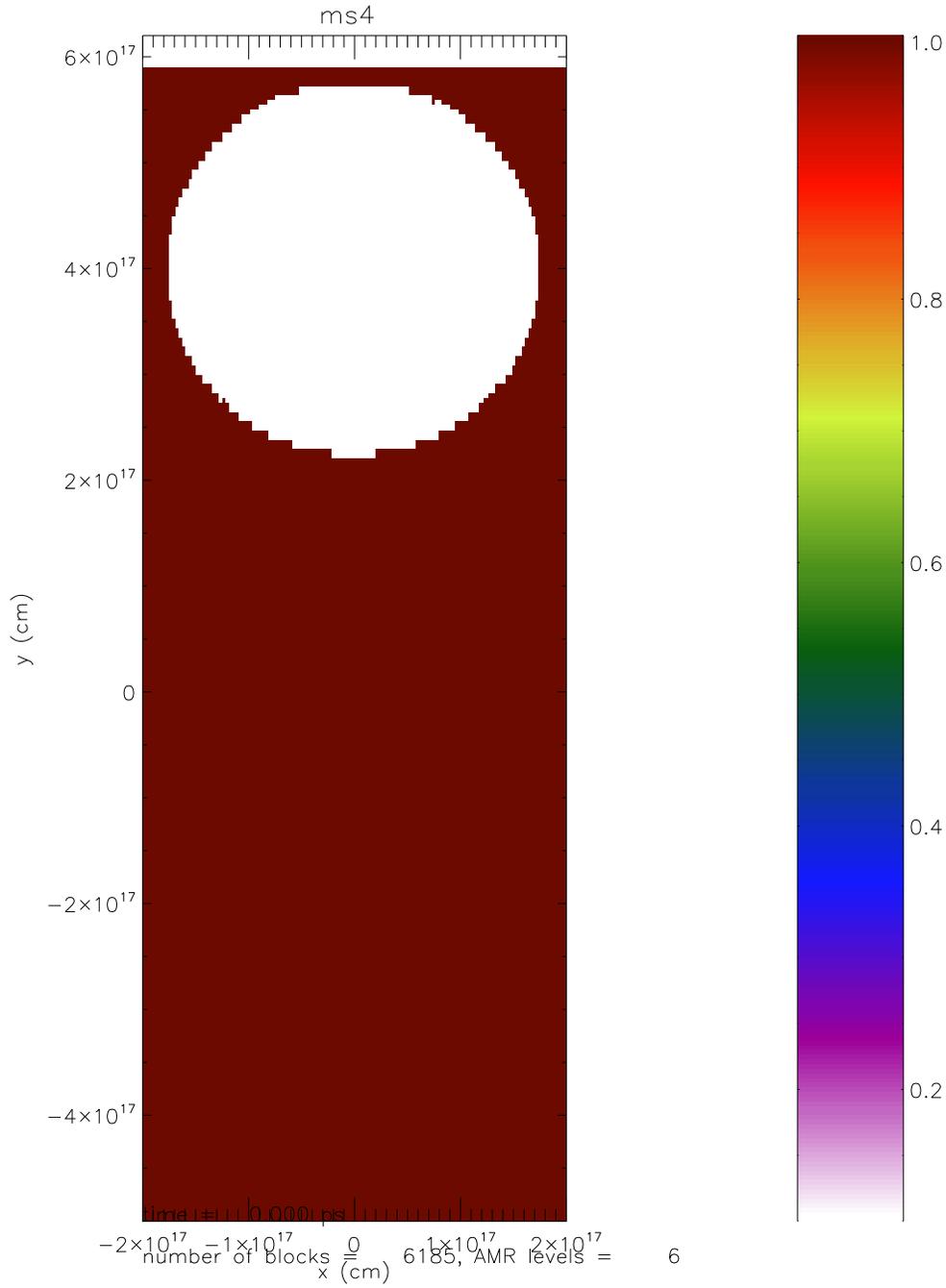}
\vspace{0.5in}
\caption{Initial cross-section ($z$ = 0) of color field 4, representing the material
initially outside the target cloud core, plotted as in Figure 1. This material is 
initially stationary.}
\end{figure}
\clearpage

\begin{figure}
\vspace{-1.0in}
\includegraphics[scale=.80,angle=0]{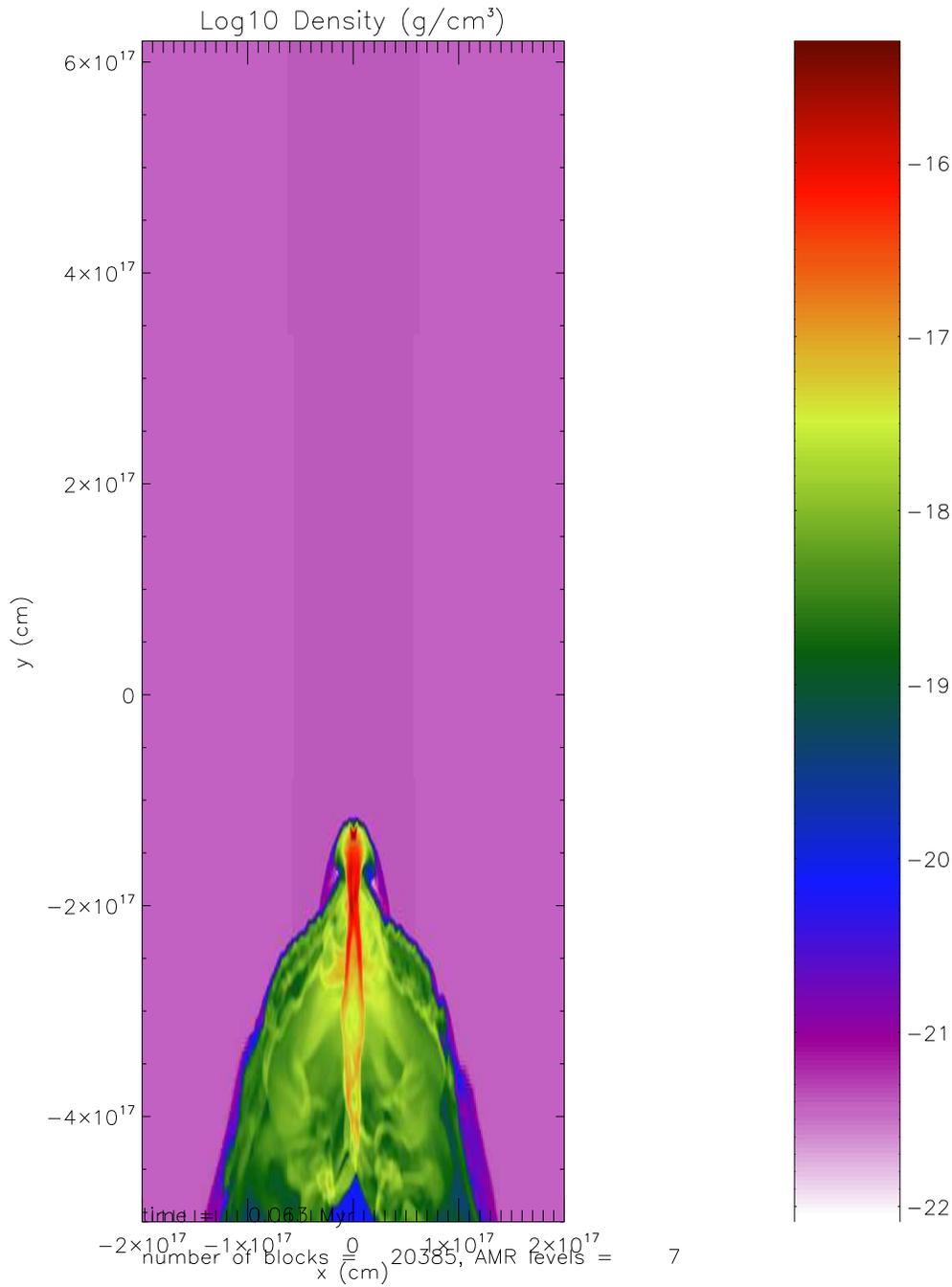}
\vspace{0.5in}
\caption{Final log density cross-section ($z$ = 0) after 0.063 Myr with seven levels 
of refinement. By this time, the portions of the shock front that did not strike the
target cloud have exited the bottom of the grid.}
\end{figure}
\clearpage

\begin{figure}
\vspace{-1.0in}
\includegraphics[scale=.80,angle=0]{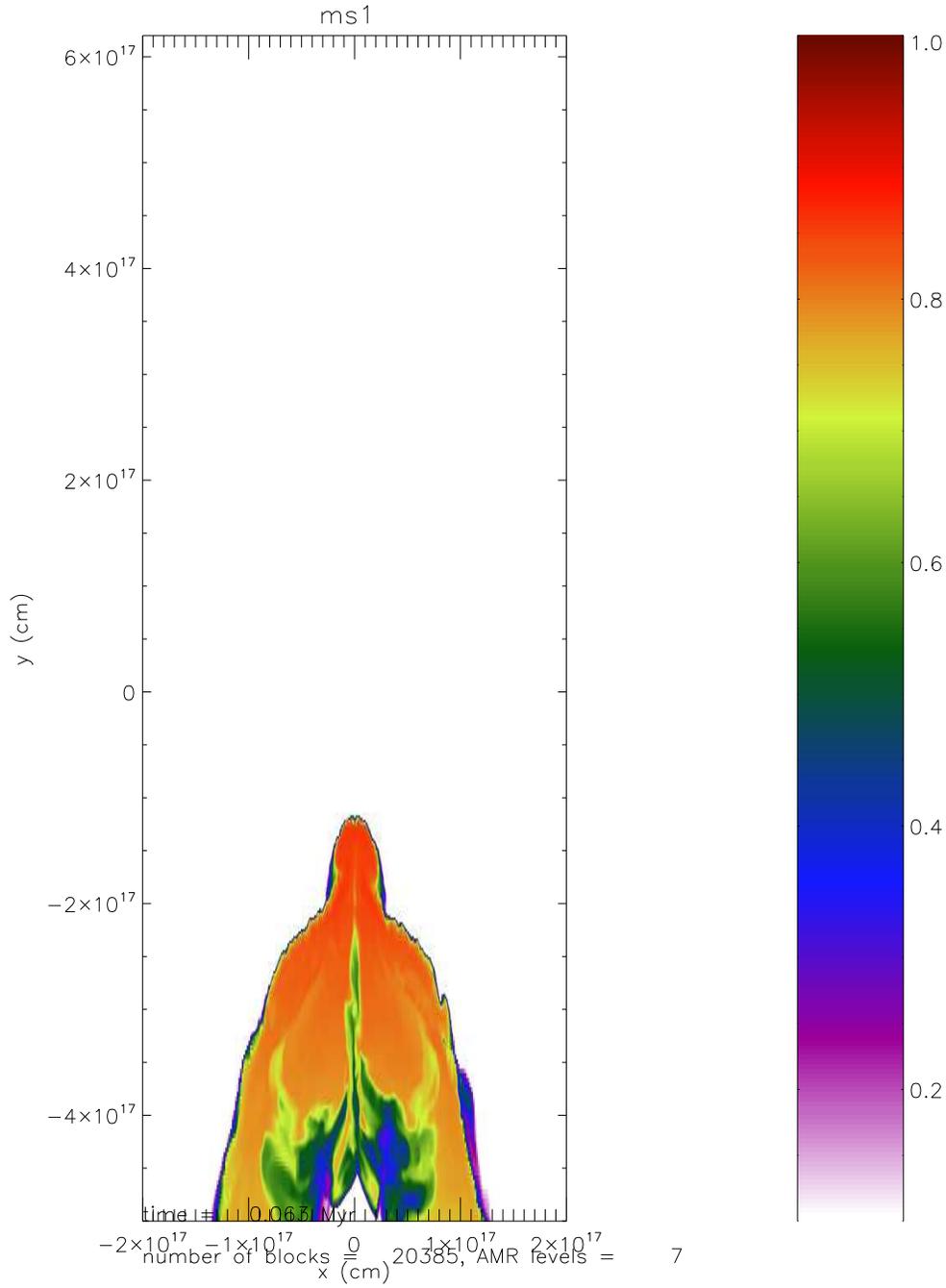}
\vspace{0.5in}
\caption{Final cross-section ($z$ = 0) of color field 1, representing the material
initially within the target cloud core, plotted as in Figure 6.}
\end{figure}
\clearpage

\begin{figure}
\vspace{-1.0in}
\includegraphics[scale=.80,angle=0]{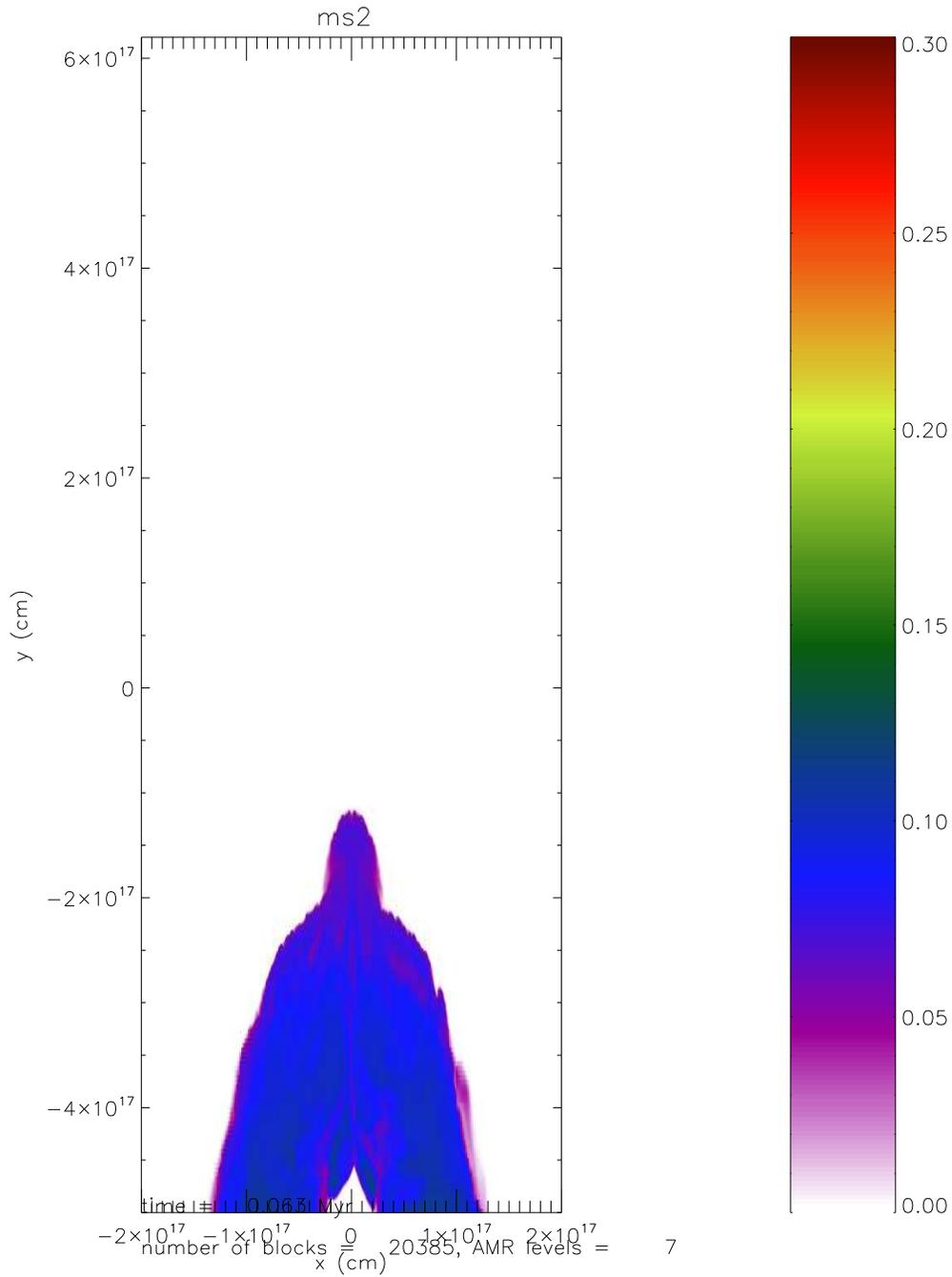}
\vspace{0.5in}
\caption{Final cross-section ($z$ = 0) of color field 2, representing the material
initially within the shock front, plotted as in Figure 6. Note the scale change
compared to Figure 3.} 
\end{figure}
\clearpage

\begin{figure}
\vspace{-1.0in}
\includegraphics[scale=.80,angle=0]{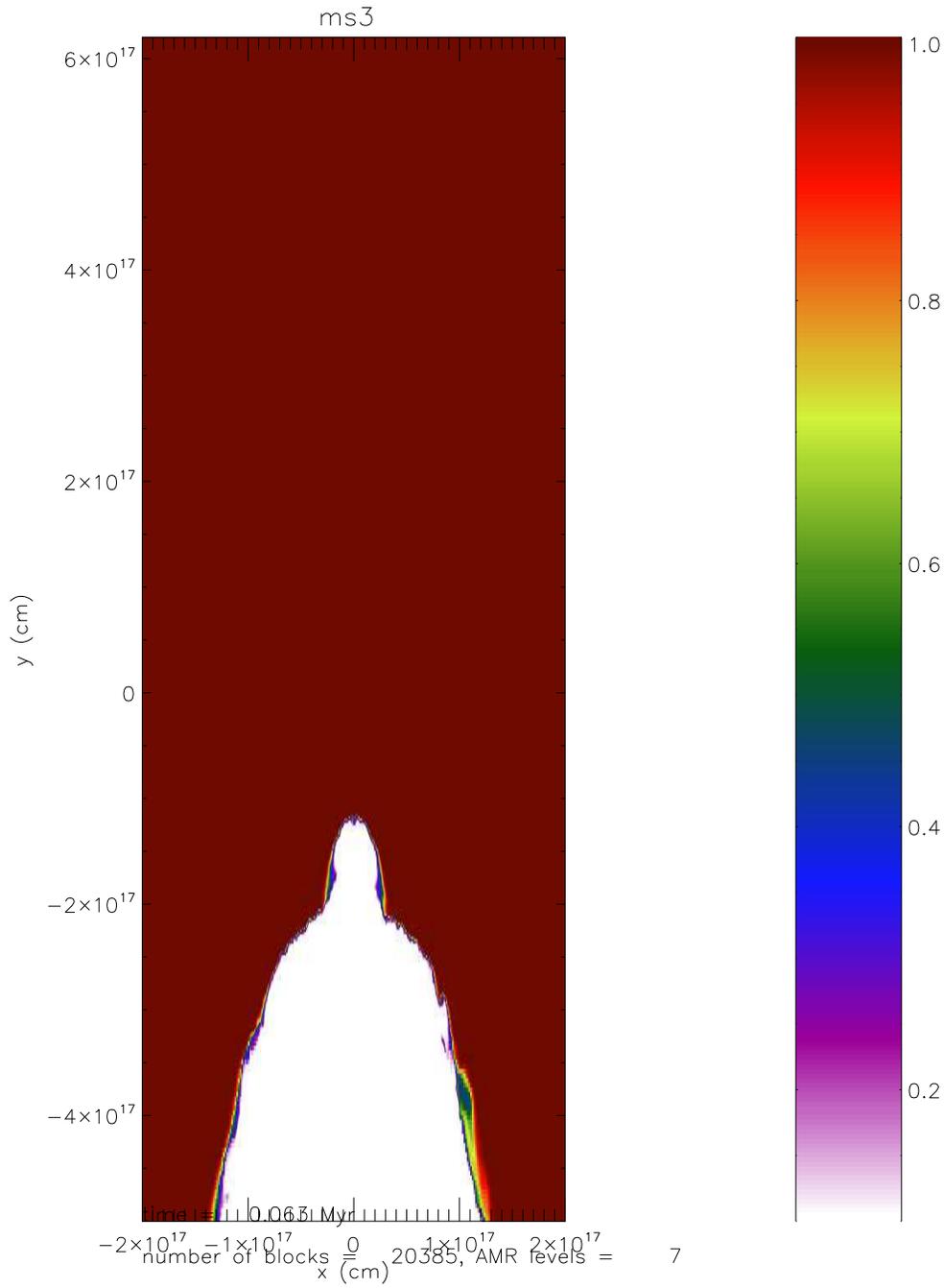}
\vspace{0.5in}
\caption{Final cross-section ($z$ = 0) of color field 3, representing the material
initially behind the shock front, plotted as in Figure 6.}
\end{figure}
\clearpage

\begin{figure}
\vspace{-1.0in}
\includegraphics[scale=.80,angle=0]{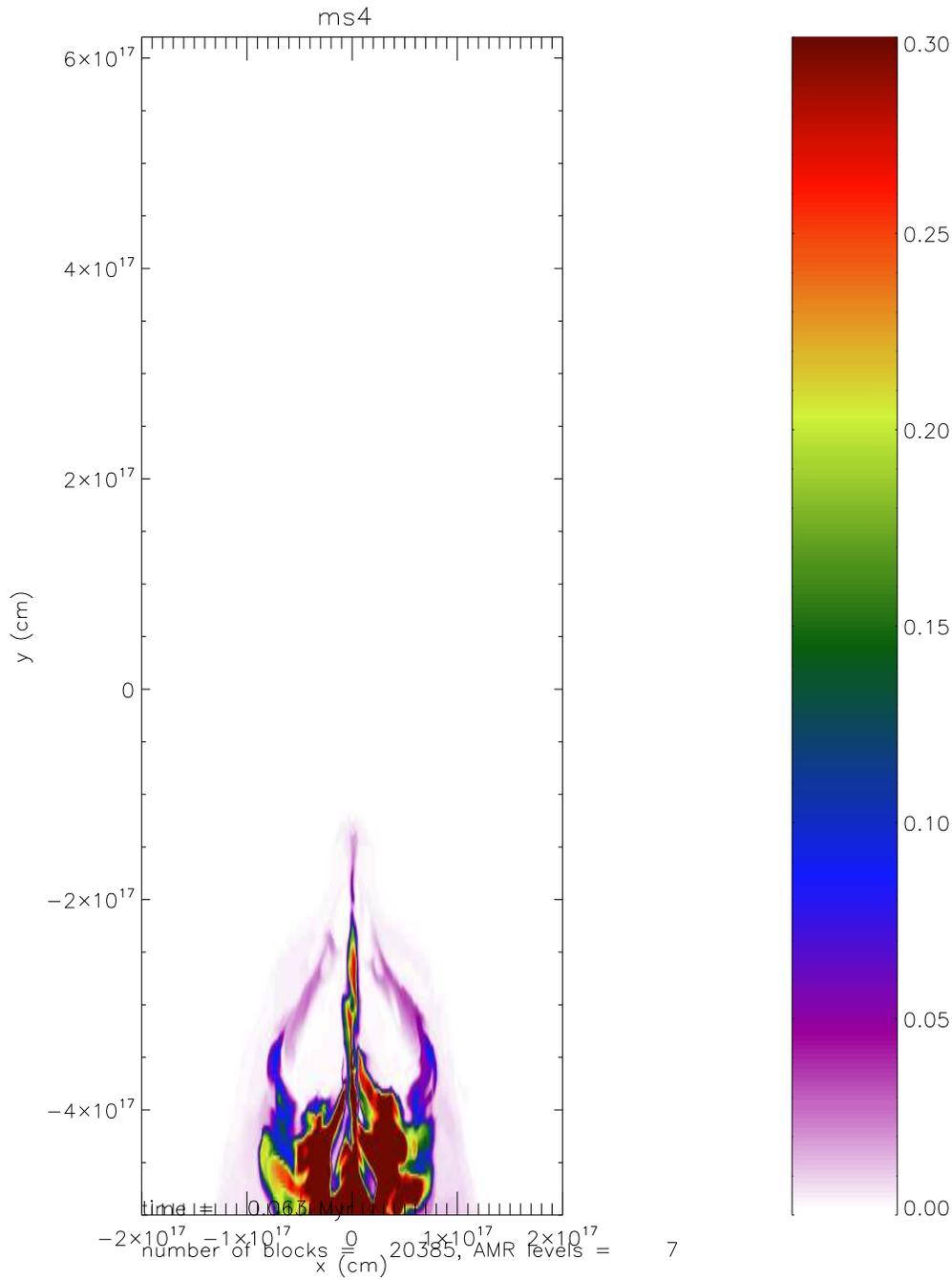}
\vspace{0.5in}
\caption{Final cross-section ($z$ = 0) of color field 4, representing the material
initially outside the target cloud core, plotted as in Figure 6. Note the scale change
compared to Figure 5.} 
\end{figure}
\clearpage

\begin{figure}
\vspace{-1.0in}
\includegraphics[scale=.80,angle=0]{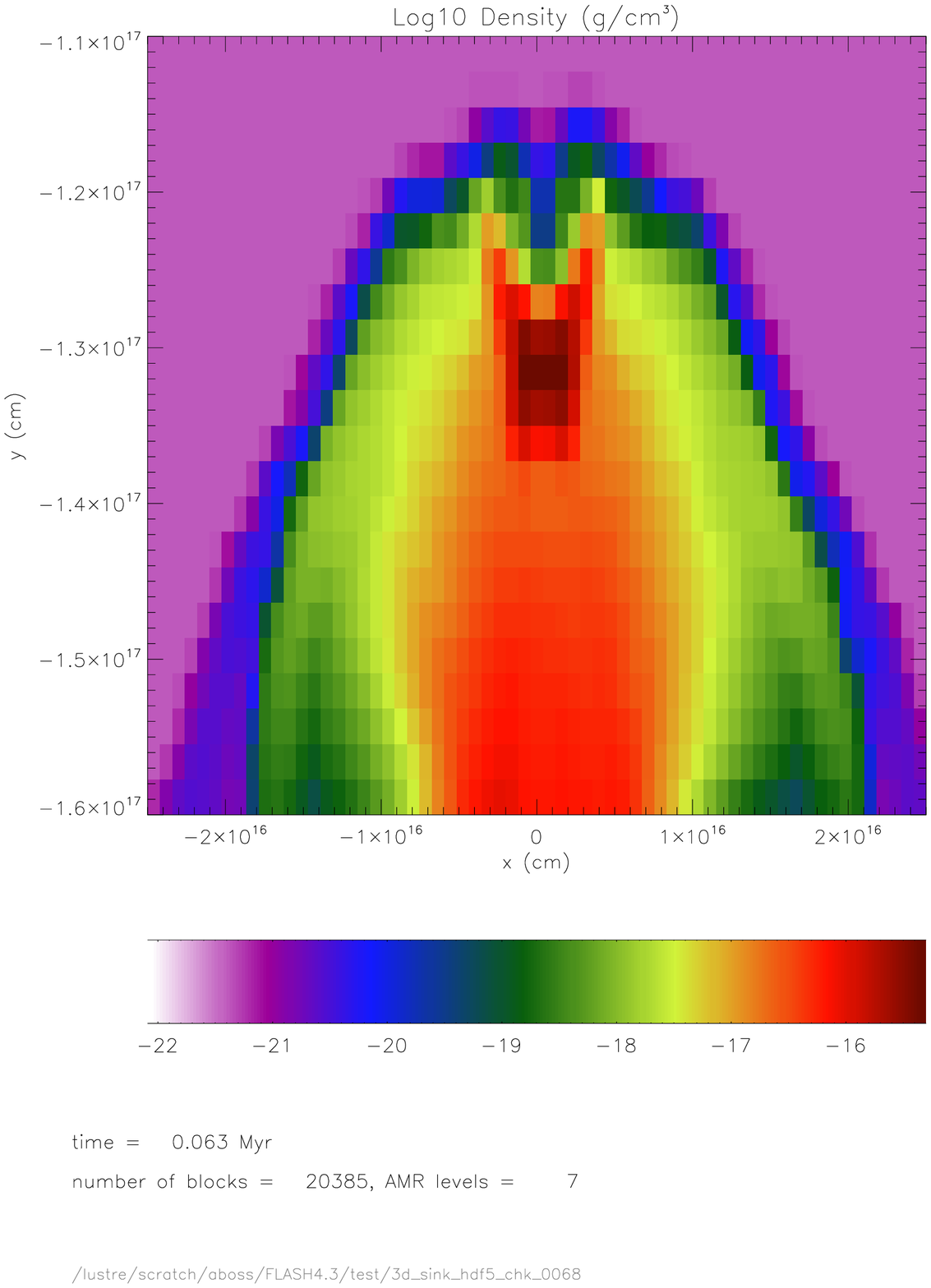}
\vspace{0.5in}
\caption{Close-up view of the final log density cross-section ($z$ = 0) after 0.063 Myr
as seen in Figure 6. The sink cell is located in the center of the density maximum and
represents the newly-formed protostar and disk system.}
\end{figure}
\clearpage

\begin{figure}
\vspace{-1.0in}
\includegraphics[scale=.80,angle=0]{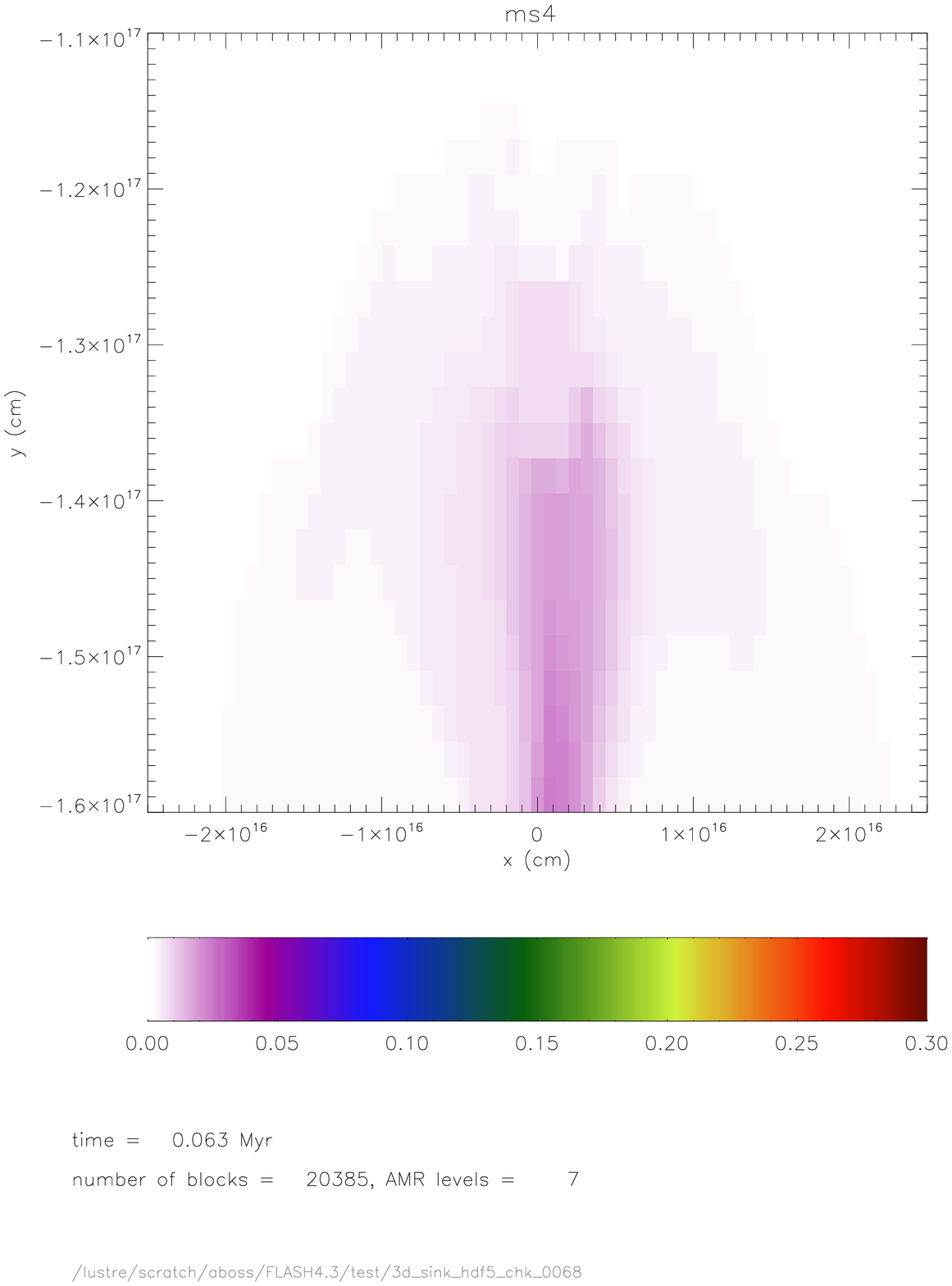}
\vspace{0.5in}
\caption{Close-up view of the final cross-section ($z$ = 0) of color field 4, 
representing the material initially outside the target cloud core, plotted as in Figure 11.}
\end{figure}
\clearpage

\begin{figure}
\vspace{-1.0in}
\includegraphics[scale=.80,angle=0]{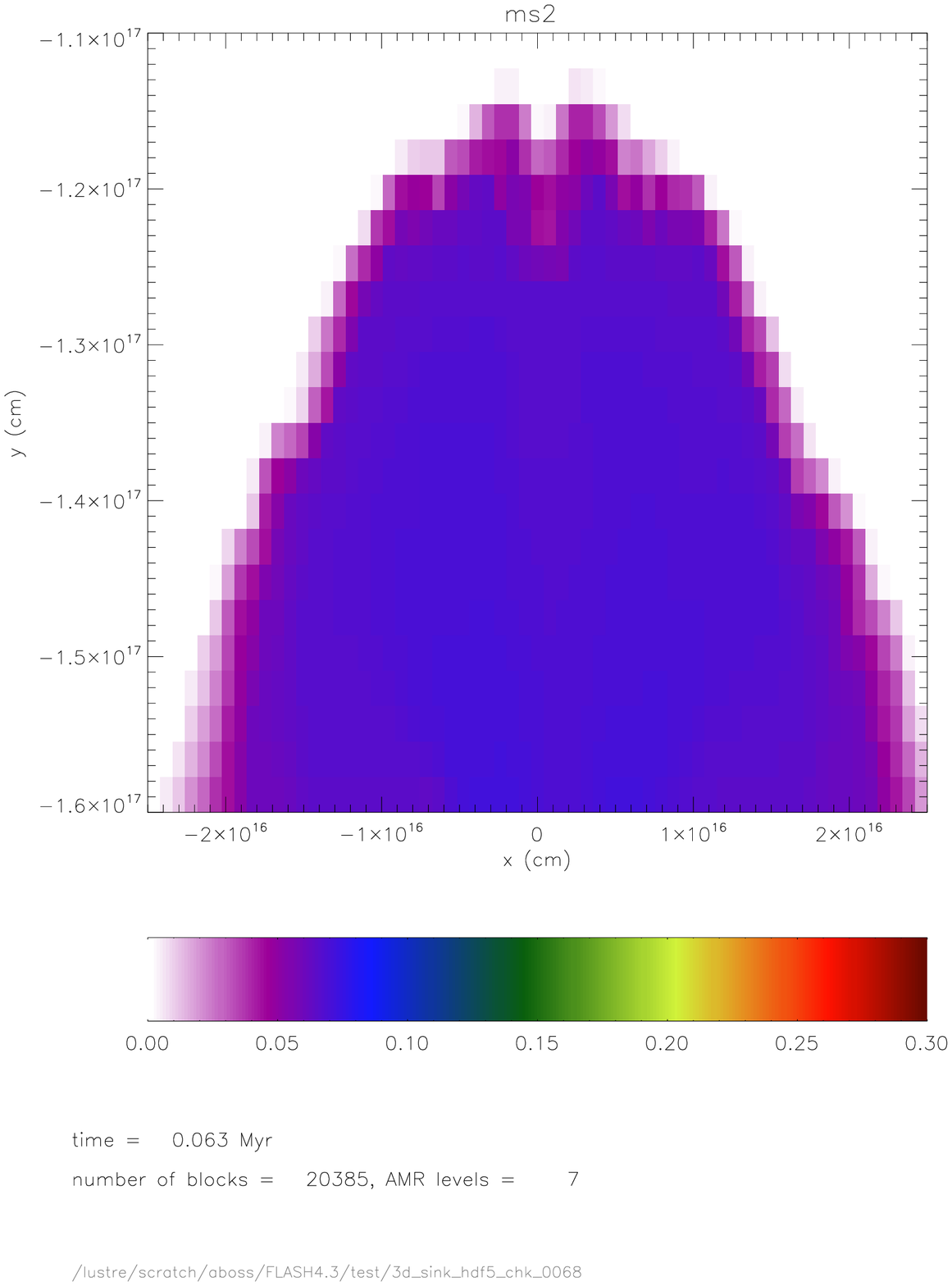}
\vspace{0.5in}
\caption{Close-up view of the final cross-section ($z$ = 0) of color field 2, 
representing the material initially inside the shock front, plotted as in Figure 11.
Note the scale change compared to Figure 3.} 
\end{figure}
\clearpage

\begin{figure}
\vspace{-1.0in}
\includegraphics[scale=.80,angle=0]{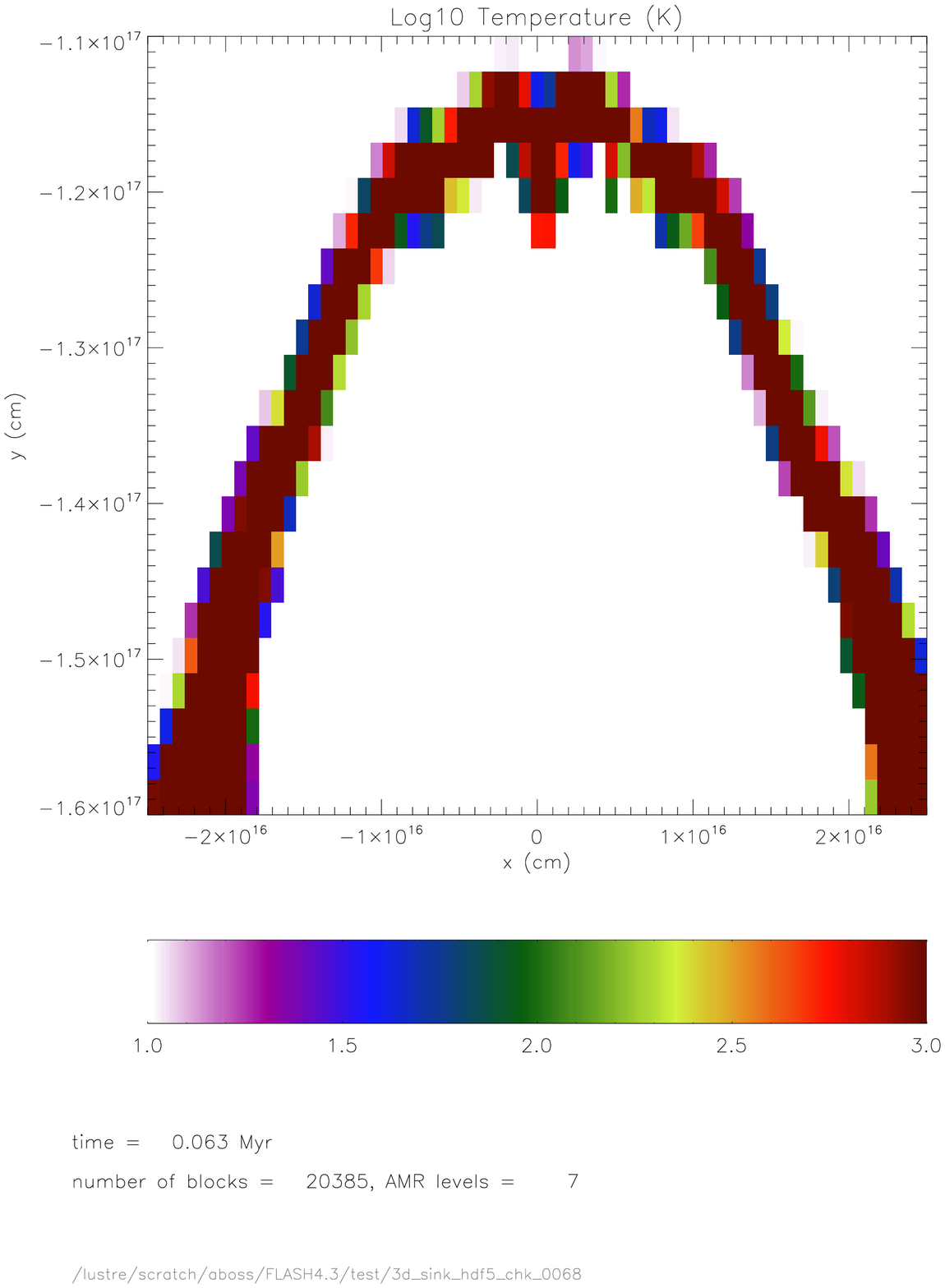}
\vspace{0.5in}
\caption{Close-up view of the final cross-section ($z$ = 0) of the log temperature
distribution, plotted as in Figure 11.}
\end{figure}
\clearpage

\end{document}